\documentclass[aps,pre,preprint,noshowpacs,floatfix,amsmath]{revtex4}
\usepackage[pdftex]{graphicx}
\usepackage{amssymb}
\usepackage{amsmath}
\usepackage[]{enumerate}
\usepackage{bm}

\DeclareBoldMathCommand{\bV}{V}
\DeclareBoldMathCommand{\bb}{b}
\DeclareBoldMathCommand{\bB}{B}
\DeclareBoldMathCommand{\bE}{E}
\DeclareBoldMathCommand{\bk}{k}

\newcommand{\dif}{\mathrm{d}}
\newcommand{\al}{\alpha}
\newcommand{\bt}{\beta}

\newcommand{\p}{\partial}

\newcommand{\mat}{Mathematica\textsuperscript{\textregistered} }
\DeclareBoldMathCommand{\bV}{V}
\DeclareBoldMathCommand{\bv}{v}
\DeclareBoldMathCommand{\bF}{F}
\DeclareBoldMathCommand{\bg}{g}
\DeclareBoldMathCommand{\bl}{\ell}
\DeclareBoldMathCommand{\bu}{u}
\DeclareBoldMathCommand{\br}{r}
\DeclareBoldMathCommand{\bx}{x}
\DeclareBoldMathCommand{\bz}{z}
\DeclareBoldMathCommand{\bg}{g}
\DeclareBoldMathCommand{\bb}{b}
\DeclareBoldMathCommand{\be}{e}
\DeclareBoldMathCommand{\bs}{s}
\DeclareBoldMathCommand{\bA}{A}
\DeclareBoldMathCommand{\bB}{B}
\DeclareBoldMathCommand{\bC}{C}
\DeclareBoldMathCommand{\bD}{D}
\DeclareBoldMathCommand{\bE}{E}
\DeclareBoldMathCommand{\bI}{I}
\DeclareBoldMathCommand{\bJ}{J}
\DeclareBoldMathCommand{\bM}{M}
\DeclareBoldMathCommand{\bL}{L}
\DeclareBoldMathCommand{\bN}{N}
\DeclareBoldMathCommand{\bP}{P}
\DeclareBoldMathCommand{\bR}{R}
\DeclareBoldMathCommand{\bS}{S}
\DeclareBoldMathCommand{\bU}{U}
\DeclareBoldMathCommand{\bW}{W}
\DeclareBoldMathCommand{\bk}{a}
\DeclareBoldMathCommand{\ba}{a}
\DeclareBoldMathCommand{\bn}{n}
\DeclareBoldMathCommand{\bp}{p}
\DeclareBoldMathCommand{\bq}{q}
\DeclareBoldMathCommand{\br}{r}

\allowdisplaybreaks
\begin{document}
\title{Symmetries of Reduced Magnetohydrodynamics}
\author{Panagiotis Koutsomitopoulos, Reese S. Lance, S. A. Yadavalli,\\ and R. D. Hazeltine}
\affiliation{Institute for Fusion Studies and Department of Physics, University of Texas at 
Austin}
\date{\today}
\begin{abstract}Lie-symmetry methods are used to determine the symmetry group of reduced magnetohydrodynamics. This group allows for arbitrary, continuous transformations of the fields themselves, along with space-time transformations. The derivation reveals, in addition to the predictable translation and rotation groups, some unexpected symmetries.  It also uncovers novel, exact nonlinear solutions to the reduced system.  A similar analysis of a related but simpler system, describing nonlinear plasma turbulence in terms of a single field, is also presented.
\end{abstract}
\maketitle

\section{Introduction} \label{sec:label}
Reduced magnetohydrodynamics \cite{Strauss_1976} (RMHD) is a simplified version of MHD, based on a combination of geometrical approximation and time-scale separation.  Most importantly, RMHD distinguishes the fast time scale of compressional Alfven waves from the slower evolution of shear-Alfven waves, and assumes that the former have relaxed to equilibrium.  This reduced fluid model was constructed in the context of magnetic-confinement fusion, and originally used in numerous studies of nonlinear tokamak plasma behaviour; an example is \cite{dahlberg1986}.  But the model has found much wider application, including solar and astrophysical research; see, for example,  \cite{zank2009, oughton2017, matt1990}.

Despite the recent interest in more complicated models---nonlinear systems that include, for example, kinetic effects (such as \cite{zoccoetal})---RMHD remains a broadly useful tool for understanding the nonlinear dynamics of magnetized plasma.  It therefore deserves a systematic study of its continuous symmetries, following the Lie-group procedure \cite{cantwell, olver}.  Thus, considering continuous transformations of RMHD's independent coordinates $(x,y,z,t)$ and dependent fields $(\psi, \phi)$, we identify those transformations which leave the form of the equations unchanged.

\section{Mathematical framework of Lie analysis}

The Lie procedure identifies symmetries of differential equations: transformations of the independent and dependent variables that leave the equations unchanged. Symmetries can reveal important properties of the analyzed system, and in some cases lead to exact nonlinear solutions. Here we briefly review Lie's recipe for symmetry analysis. Thorough discussions can be found in textbooks \cite{cantwell, olver}.

Consider the a differential equation $\Xi[x^i, u^j, u^k_{x^\alpha x^\beta ...}]  =0$ where $x^i$ are independent variables, $u^j$ are dependent variables and $u^k_{x^\alpha x^\beta ...}$ are their derivatives w.r.t. $x^\alpha,x^\beta$ and so on. We consider smooth transformation functions $(X^i,U^j)$ of the all the variables $(x^i,u^j)$ respectively, parameterized by the single real parameter $\epsilon$ such that $X^i(x^i,u^j;\epsilon=0)=x^i$ and $U^j(x^i,u^j;\epsilon=0)=u^{j}$. The set of such transformations will be the Lie group for this differential equation.

We can generate a vector field by looking at ``generators:'' infinitesimal changes around $\epsilon = 0$. This vector field, $\mathcal{L}$, is defined component-wise as
    \begin{equation*}
        \mathcal{L}\ =\ \left(\frac{dX^i}{d\epsilon}\Big|_{\epsilon=0},\frac{dU^j}{d\epsilon}\Big|_{\epsilon=0} \right)\ =:\ \left(\xi^i,\eta^j\right)
    \end{equation*}
If the differential equation in question involves only $(x^i,u^j)$ (no derivatives), the Lie operator that perturbs the differential equation is just this vector field:
\begin{equation*}
    \mathcal{L}\ =\ \xi^i \partial_{x^i} +\eta^j \partial_{u^j}
\end{equation*}

If there are derivative terms in our differential equation, such as $u^k_{x^\alpha x^\beta ...}$, we must \textit{prolong} the Lie operator to account for variations in these derivative terms too. We label the generators associated with such terms as $\eta^k_{\{x^\alpha x^\beta ...\}}$. Thus we define the prolonged Lie operator, $\mathcal{L}^*$, as
\begin{equation*}
\mathcal{L}\rightarrow\mathcal{L}^*=\xi^i \partial_{x^i} + \eta^j \partial_{u^j} + \eta^k_{\{x^\alpha x^\beta ...\}}\partial_{u^k_{x^\alpha x^\beta ...}}
\end{equation*}
The Lie procedure expresses $\eta^k_{\{x^\alpha x^\beta ...\}}$  as derivatives of  $\xi^i$ and $\eta^j$ \cite{cantwell}. Acting this prolonged Lie operator upon a differential equation is equivalent to looking at a first-order change to a differential equation when we consider infinitesimal changes to the associated variables of all derivative orders.

By definition, the continuous symmetries of a differential equation are those transformations that do \textit{not} perturb the differential equation. Thus the symmetry condition for $\Xi$ is,
\begin{equation*}
\mathcal{L}^* \Xi = 0
\end{equation*}
By computing and simplifying this condition, we obtain what are known as the ``determining equations'' for the continuous symmetries for $\Xi$. Solution of these equations provides the continuous symmetries in the system. Derivation of the determining equations is straightforward but lengthy; fortunately a number of software packages that perform the derivation are available.  For this paper we have used a combination of ``hands-on'' analysis and a \mat package provided by Cantwell \cite{cantwell}.

\section{CHM symmetry analysis}
\subsection{Differential equation of CHM}
To begin with a relatively simple example, we apply the Lie symmetry analysis procedure to the Charney-Hasegawa-Mima equation (CHM) \cite{CHM}. This nonlinear third-order partial differential equation was constructed to describe plasma turbulence. It uses a single dependent variable, the electrostatic potential $\phi$, and is given by
\begin{equation}
    \partial_t U + [\phi, U] = \partial_t \phi \label{chm}
\end{equation}
Here $U \equiv \nabla^2_\perp\phi = \phi_{xx} + \phi_{yy}$, is the plasma vorticity. Note that partial derivatives are indicated by subscripts. The bracket is defined by $$[\phi,U] \equiv \phi_x U_y - \phi_y U_x = \phi_x(\phi_{xxy}+\phi_{yyy}) - \phi_y(\phi_{xxx}+\phi_{yyx})$$

The symmetries of a related nonlinear system have been analyzed previously \cite{HK2008}.
    
\subsection{CHM determining equations} \label{de}
Beginning with the differential operator
\[
\mathcal{L} = \xi^{x}\p_{x}  + \xi^{y}\p_{y} +  \xi^{z}\p_{z} +  \xi^{t}\p_{t} + \eta \p_{\phi}
\]
we require the CHM equation to be invariant under the action of the prolonged operator $\mathcal{L}^{*}$, as described above.  This requirement leads to the following determining equations:
\begin{align}
\nabla_{\perp}^{2}\eta_{t}  - \eta_{t} &= 0 \label{d1}\\
\nabla_{\perp}^{2}\eta_{\phi}  - 2 \xi_{x}^{x} &= 0 \label{d2} \\
\eta_{t\phi} - 2 \xi^{x}_{tx} &= 0 \label{d3}\\
\eta_{\phi} + \xi^{t}_{t} -\xi^{x}_{x}  - \xi^{y}_{y} &= 0 \label{d4}\\
\xi^{y}_{x} + \xi^{x}_{y} &= 0 \label{d5}\\
\xi^{x}_{x} - \xi ^{y}_{y} &= 0 \label{d6}\\
\eta_{\phi xx} - \eta_{\phi yy}&= 0 \label{d7}\\
\nabla_{\perp}^{2}\eta_{y} +\xi_{t}^{x} &= 0 \label{d8}\\
\nabla_{\perp}^{2}\eta_{x}  - \xi^{y}_{t} &= 0 \label{d9}\\
\eta_{x\phi} - 2\xi^{y}_{xy} &= 0 \label{d10}\\
\eta_{y \phi} - 2\xi^{x}_{xy} &= 0 \label{d11}\\
2\eta_{\phi x} - \nabla_{\perp}^{2}\xi^{x}&= 0\label{d12}\\
2 \eta_{\phi y} - \nabla_{\perp}^{2}\xi^{y}  &= 0 \label{d13}\\
\eta_{x} - \xi_{t}^{y} &= 0 \label{d14}\\
\eta_{y} + \xi^{x}_{t} &= 0 \label{d15}\\
\eta_{\phi \phi} &= 0 \label{d16} \\
\xi^{t}_{\phi} = \xi^{x}_{\phi} =\xi^{y}_{\phi} = \xi^{z}_{\phi} &= 0 \label{d17}\\
\xi^{t}_{x} = \xi^{t}_{y} &= 0 \label{d18}\\
\xi^{z}_{t} = \xi^{z}_{x} =\xi^{z}_{y}  &= 0 \label{d19}
\end{align}

Solution of the determining equations is straightforward; we sketch the procedure here. First observe that (\ref{d5}) and (\ref{d6}) imply
\begin{align*}
    \xi^x\ &=\ a(z,t) + b(z,t)y + C_{x}(z,t)\\
    \xi^y\ &=\ a(z,t)y - b(z,t) x + C_{y}(z,t)
\end{align*}
where $a$, $b$ and the $C_{j}$ are arbitrary functions. Next we notice from (\ref{d16}) that $\eta$ is linear in $\phi$,
$$
    \eta = \eta_0(x,y,z,t) + \eta_1(x,y,z,t)\phi
$$
while (\ref{d2}) implies that
$$
    \nabla_\perp^2\eta_1 = 2a
$$
Noting that the second terms in both (\ref{d12}) and (\ref{d13}) vanish, we can conclude
$$
    \eta_{1x}\ =\ \eta_{1y}\ =\ 0
$$
This implies that $a = 0$, whence (\ref{d3}) implies
$$
    \eta_{1t}\ =\  0
$$
Working through the remaining equations in a similar manner, we are led to conclude
\begin{align*}
    \eta\ &=\ \eta_0(z) + \eta_1(z) \phi\\
    \xi^x\ &=\ b(z)y+C_x(z)\\
    \xi^y\ &=\ -b(z)x + C_y(z)\\
    \xi^z\ &\equiv\ \xi^z(z)\\
    \xi^t\ &=\ -\eta_1(z)  t+ C_t(z)
\end{align*}
The functions $b(z),\ \eta_0(z)$, $\xi^z(z)$, $\eta_{i}$ and $C_{x,y}$ are arbitrary. Note that the terms involving $b(z)$ describe a $z$-dependent rotation in the $(x,y)$ plane.

\subsection{Lie symmetries of CHM}
We have found the following (unsurprising) symmetries of the CHM model:
\begin{enumerate}
    \item The $x$ and $y$ origins can be displaced, by amounts varying in $z$. 
    \item The coordinates may be rotated about the $z$ axis, also by amounts varying in $z$. 
    \item The $z$ origin can be displaced. 
    \item $\phi$ can be scaled by a factor $\lambda$, provided there is an accompanied ``inverse'' scale of $t$ in the following sense: 
    \begin{align*}
        \phi  \rightarrow \lambda(z)\phi\\
        t \rightarrow \ \frac{1}{\lambda(z)}t
    \end{align*}
    Note that the scale factor $\lambda$ can vary with $z$.
    \item $\phi$ can be translated by a function which depends only on $z$. 
\end{enumerate}
The direct verification of these symmetries is straightforward.

\subsection{Exact solutions of CHM}
We can use these symmetries to generate families of exact solutions for $\phi$. We begin with an exact solution that we can transform---using the symmetries---to produce such a family. For CHM, these results are not very interesting, but this is a primer for the following RMHD analyses.

In CHM, we consider the class of $\phi$ solutions which are cylindrically symmetric about the $z$ axis. In this case, it is convenient to work in cylindrical coordinates, with $r = \sqrt{x^{2} + y^{2}}$. Assuming a  separable solution
\[
\phi = f(z,t)g(r)
\]
we find that $g$ can be chosen to be a modified Bessel function,
$$
    K_{0}(r) = \int_0^\infty e^{-r\hspace{.5mm}\text{cosh}\hspace{.5mm} s}\hspace{.5mm} \dif s
$$
Thus CHM has the exact, cylindrically symmetric solution
\begin{equation}
    \phi=K_0(r) f(z,t)
\end{equation}
where the function $f(z,t)$ is arbitrary and we have ignored a solution growing exponentially with $r$. 

The only other interesting symmetries in this system are the $\phi-t$ scaling symmetry and $\phi$ translation symmetry. We use that to observe the transformations
\begin{equation}
    \Tilde{\phi}=\lambda(z) \phi + A(z);\ \Tilde{t}=t/\lambda(z)
\end{equation}
where $\lambda(z)$ and $A(z)$ are arbitrary functions. By re-substituting we find, suppressing tildes,
\begin{equation}
   \phi=\lambda(z) K_{0}(r)f(z, \lambda t)  + A(z)
\end{equation}
is also a (slightly non-trivial) family of exact solutions with freedom in $\lambda$ and $A$. 

\section{RMHD symmetry analysis}

\subsection{Differential equations of RMHD}

We now apply the Lie symmetry analysis to a more complicated system, which yields more interesting results. The RMHD system is a set of two partial differential equations of  third order that involve space and time derivatives, and  two fields, $\psi$ and $\phi$. Here $\psi$ is a normalized measure of the longitudinal vector potential, $A_{z}$, and $\psi$ measure the electrostatic potential. The plasma current is denoted by $J =  \nabla^{2}_{\perp} \psi$ and the vorticity by $U = \nabla^{2}_{\perp} \phi$ (as in CHM). Our analysis is applied to the original, simplest version of RMHD, as given by Strauss  \cite{Strauss_1976}, to which the reader is referred for physical interpretation of the model.  Thus we have

    \begin{equation*}
         \partial_t U + [\phi, U] + \nabla_{\parallel} J = 0
    \end{equation*}
    \begin{equation*}
        \partial_t \psi + \nabla_{\parallel} \phi = 0
    \end{equation*}

where
    \begin{align*}
        J = \nabla^{2}_{\perp} \psi \\
        \nabla^{2}_{\perp} f \equiv \partial^2_x f + \partial^2_y f \\
        \nabla_{\parallel} f \equiv \partial_z f - [\psi,f] \\
    \end{align*}
    
The Lie operator for the RMHD equation is defined as 

\begin{gather*}
    \mathcal{L} \equiv \xi^x \partial_x + \xi^y \partial_y + \xi^z \partial_z + \xi^t \partial_t + \eta^\psi \partial_\psi + \eta^\phi \partial_\phi
\end{gather*}

\subsection{RMHD determining equations}

As in the CHM analysis, this operator must be prolonged to allow its operation on the various derivatives.  The explicit form of the prolonged operative, which involves many terms, is omitted here.  Instead we turn our attention to the determining equations, given by 

\begin{equation}
    \xi^x_\phi = \xi^y_\phi = \xi^z_\phi = \xi^t_\phi = 0
\end{equation}
\begin{equation}
    \xi^x_\psi = \xi^y_\psi = \xi^z_\psi = \xi^t_\psi = 0
\end{equation}
\begin{equation}
   \xi^t_x = \xi^z_x = 0
\end{equation}
\begin{equation}
    \xi^t_y = \xi^z_y = 0
\end{equation}
\begin{equation}
    \xi^t_z = 0
\end{equation}
\begin{equation}
    \xi^z_t = 0
\end{equation}
\begin{equation}
    \eta^{\psi}_\phi = 0
\end{equation}
\begin{equation}
    \eta^{\phi}_\psi = 0
\end{equation}
\begin{equation}
    \eta^{\psi}_{\psi\psi} = 0
\end{equation}
\begin{equation}
    \eta^{\phi}_{\phi\phi} = 0
\end{equation}
\begin{equation}
    \xi^x_y + \xi^y_x = 0
\end{equation}
\begin{equation}
    \xi^y_y - \xi^x_x = 0
\end{equation}
\begin{equation}
    \xi^x_z - \eta^{\psi}_y = 0
\end{equation}
\begin{equation}
    \xi^y_z + \eta^{\psi}_x = 0
\end{equation}
\begin{equation}
    \xi^x_t + \eta^{\phi}_y = 0
\end{equation}
\begin{equation}
    \xi^y_t - \eta^{\phi}_x = 0
\end{equation}
\begin{equation}
    \eta^{\psi}_{\psi y} - 2\xi^x_{xy} = 0
\end{equation}
\begin{equation}
    \eta^{\psi}_{\psi x} - 2\xi^y_{xy} = 0
\end{equation}
\begin{equation}
    \eta^{\phi}_{\phi y} - 2\xi^x_{xy} = 0
\end{equation}
\begin{equation}
    \eta^{\phi}_{\phi x} - 2\xi^y_{xy} = 0
\end{equation}
\begin{equation}
    \eta^{\psi}_{\psi z} = 2\xi^x_{xz}
\end{equation}
\begin{equation}
    \eta^{\psi}_{\psi z} = 2\xi^y_{yz}
\end{equation}
\begin{equation}
    \eta^{\phi}_{\phi t} = 2\xi^x_{xt}
\end{equation}
\begin{equation}
    \eta^{\phi}_{\phi t} = 2\xi^y_{yt}
\end{equation}
\begin{equation}
    \xi^x_z - \eta^{\psi}_y = 0
\end{equation}
\begin{equation}
    \xi^y_z + \eta^{\psi}_x = 0
\end{equation}
\begin{equation}
    \xi^x_t + \eta^{\phi}_y = 0
\end{equation}
\begin{equation}
    \xi^y_t - \eta^{\phi}_x = 0
\end{equation}
\begin{equation}
    \eta^{\psi}_t + \eta^{\phi}_z = 0
\end{equation}
\begin{equation}
    \eta^{\phi}_\phi - \xi^y_y - \xi^x_x +\xi^t_t = 0
\end{equation}
\begin{equation}
    \eta^{\phi}_\phi - \eta^{\psi}_\psi + \xi^z_z - \xi^t_t = 0
\end{equation}
\begin{equation}
    \eta^{\psi}_{zyy} + \eta^{\psi}_{zxx} + \eta^{\phi}_{txx} + \eta^{\phi}_{tyy} = 0
\end{equation}
\begin{equation}
    \eta^{\phi}_\phi - 2 \eta^{\psi}_\psi + \xi^y_y + \xi^x_x - \xi^t_t = 0
\end{equation}
Analysis on the determining equations, as in subsection \ref{de}, leads to the following conclusions regarding the generators $\xi^{i}$ and $\eta^{i}$:

    \begin{equation}
        \xi^{t}(t) = \delta t + \chi^1 \label{g1}
    \end{equation}
    \begin{equation}
        \xi^{z}(z) = \delta z + \chi^2  \label{g2}
    \end{equation}
    \begin{equation}
        \xi^{x}(x, y, z, t) = \frac{\kappa + \delta}{2} x + \beta (z, t) y + R (z, t)  \label{g3}
    \end{equation}
    \begin{equation}
        \xi^{y}(x, y, z, t) = -\beta (z, t) x + \frac{\kappa + \delta}{2} y + S (z, t)   \label{g4}
    \end{equation}

    \begin{equation} \begin{split}
        \eta^{\psi}(x, y, z, t) &= \kappa \psi + \frac{1}{2} r^2 \beta_z(z,t) - S_z(z,t) [x + \frac{1}{2} R(z,t)]\\ &+ R_z(z,t) [y + \frac{1}{2} S(z,t)] - F_z(z,t) \end{split}   \label{g5}
    \end{equation}
    \begin{equation}\begin{split}
        \eta^{\phi}(x, y, z, t) &= \kappa \phi - \frac{1}{2} r^2 \beta_t(z,t) + S_t(z,t) [x + \frac{1}{2} R(z,t)] \\ &- R_t(z,t) [y + \frac{1}{2} S(z,t)] + F_t(z,t) \end{split}  \label{g6}
    \end{equation}
    
Here $\kappa$, $\delta$, and $\chi^i$ are constants, while $\beta$, $F$, $R$, and $S$ are functions of $z$ and $t$. The function $\beta$ must be a solution to the wave equation
\begin{equation}
        \beta_{zz} - \beta_{tt} = 0   \label{g7}
\end{equation}
which, under RMHD normalizations, describes the shear-Alfv\'en  wave.

\subsection{Lie symmetries of RMHD} \label{disc}
We have found the following exact symmetries of RMHD:

\begin{enumerate}
\item Coordinate translations:
We can translate each variable $(x,y,z,t)$ by arbitrary fixed amounts, corres\-ponding to $\chi^{1,2}$ and \emph{constant} values for $(R,S)$.
\item Coordinate rotations:
We can rotate in the transverse $(x,y)$-plane by arbitrary fixed angles, corresponding to a constant value for $\bt$.  When $\bt$ is not constant, the rotations require simultaneous transformation of the fields, discussed below.
\item Dilations:
There are two types of dilation symmetries. \begin{enumerate}[(i)] \item When all parameters and functions vanish except $\delta$, we have dilation in $z$ and $t$, simultaneous with ``half-strength'' dilation in $x$ and $y$.  \item When only $\kappa$ does not vanish, we dilate simultaneously in $(x,y,\psi, \phi)$. \end{enumerate}
\item Gauge transformation:
The function $F(z,t)$ yields a conventional gauge transformation, involving only $z$ and $t$, as noted in previous work \cite{WHL2018}.  The transverse coordinates do not appear because the RMHD model does not include a perpendicular vector potential.

 \item We have found an ``Alfv\'enic'' gauge transformation, corresponding to non-constant $\bt$. It is a gauge transformation with regard to the variables $z$ and $t$, and it necessarily propagates at the Alfv\'en speed. Thus the general RMHD gauge transformation uses the function
\[
G(x,y,z,t) = \frac{r^{2}}{2}\bt(z,t) - F(z,t)
\]
and the gauge transformation
\[
\psi \rightarrow \psi + G_{z}, \,\,\,  \phi \rightarrow \phi - G_{t}
\]
is an exact symmetry.  We call this transformation Alfv\'enic because the function $\bt$ must satisfy the wave equation
\[
\bt_{zz} = \bt_{tt}
\]
which is the RMHD-normalized version of the Alfv\'en wave equation.  Note that the transformation is necessarily accompanied by a coordinate rotation in the transverse plane.  Thus the symmetry leads to nonlinear, helically twisted Alfv\'en waves, as exact solutions to RMHD.  A version of this symmetry was found previously \cite{WHL2018}.
\item We have found a peculiar and novel translation of the coordinates and fields,
\begin{eqnarray*}
x &\rightarrow& x + R, \\
 y &\rightarrow& y + S, \\
\psi &\rightarrow& \psi + R_{z}\left(y + \frac{1}{2}S\right) - S_{z}\left(x + \frac{1}{2}R\right), \\
\phi &\rightarrow& \phi - R_{t}\left(y + \frac{1}{2}S\right) + S_{t}\left(x + \frac{1}{2}R\right)
\end{eqnarray*}
where $R$ and $S$ are arbitrary functions of $z$ and $t$.  Notice that this transformation, while it does not affect the plasma current or vorticity, is fully nonlinear, involving the bracket. In the special case $R = \al S$, where $\al$ is a constant, the transformation becomes a gauge transformation. 
\end{enumerate}

\subsection{Exact Solutions for RMHD} \label{exact}

Because RMHD has a null solution $(\phi = 0 = \psi)$, any symmetry involving a translation of the dependent variables yields an exact nonlinear solution. More generally, suppose we have a general translation symmetry for a dependent variable $A$:
     \begin{equation*}
        \tilde{A} = A + \alpha
    \end{equation*}
     The condition imposed on all symmetric solutions requires that both $\tilde{A}$ and $A$ are solutions of RMHD. A trivial solution can be picked for $A$, yielding the exact solution $ \tilde{A}$. In RMHD  exact solutions can be found from the symmetries related to $R$ and $S$, $F$ and $\beta$.

    For the symmetry involving $F$, we obtain the easily verified solution
    \[
    \psi = -F_z(z, t), \phi = F_t(z, t)
    \]
    for any function $F(z, t)$.  
    
    For the symmetry involving $\beta$ we find 
    \[
    \psi = \frac{1}{2} r^2 \beta_z,\,\,\, \phi = -\frac{1}{2} r^2 \beta_t
    \]
  where $r^2 = x^2 +y^2$ and $\beta$ is a solution to the wave equation. Since the full symmetry transformation requires a rotation in the transverse plane, the Alfv\'en wave twists as it propagates.
        
    For the symmetry involving $R$ and $S$, both dependent and independent variables are transformed.  Distinguishing the transformed quantities by tildes and transforming the null solution, we have
  \begin{eqnarray*}
  \tilde{\psi}(x,y)  & = &-S_{z}(x + R/2) + R_{z}(y + S/2)  \\
\tilde{\phi}(x,y)  & = & S_{t}(x + R/2) -  R_{t}(y + S/2)
\end{eqnarray*}
and 
\[
\tilde{x} = x + R, \,\,\, \tilde{y} = y+ S
\]
After expressing the transformed fields in terms of the transformed coordinates and suppressing all tildes, we obtain the exact nonlinear solution
\begin{eqnarray}
\psi(x,y) &=& -S_{z}(x - R/2) + R_{z}(y - S/2) \\
\phi(x,y) &=& S_{t}(x - R/2) - R_{t}(y - S/2)    
\end{eqnarray}
   for any functions $R(z,t)$ and $S(z,t)$.  To verify this solution explicitly it suffices to note that
\[
\psi_{t} + \phi_{z} = R_{t}S_{z}  - R_{z}S_{t}  = [\psi,\phi]
\]
The solution is fully nonlinear, crucially involving the bracket.
      
 Of course additional exact solutions can be generated by combining the various transformations.       
        
\section{Summary}
The main conclusion of this work is given by (\ref{g1})--(\ref{g7}), giving the generators of the Lie symmetry group of RMHD. A qualitative discussion of these transformations is given in subsection \ref{disc}.  Some of these symmetries, such as gauge symmetry, are not surprising, but others have unexpected form. Aside from such intrinsic interest, the Lie symmetries could be useful in verifying numerical implementations RMHD, as well as aiding the interpretation of numerical results. 

In subsection \ref{exact}, the Lie symmetries were used to construct exact nonlinear solutions to RMHD; some of these solutions display novel features.

We have also analyzed, primarily to exhibit the procedure in a relatively simple case, the nonlinear fluid model CHM.  In this case the Lie methodology produced symmetries that might have been anticipated.

\section*{Acknowledgements}
We thank Ryan White and Volker Bromm for helpful advice. Two of us (R.S.L. and S.A.Y.) wish to thank everyone in and around the Kodosky Reading Room at the University of Texas for their continuous support. This work was supported by the Department of Physics, University of Texas at Austin, and by the US Department of Energy, Grant No. DOE ER54742.


\begin{thebibliography}{10}

\bibitem{Strauss_1976}
H.R. Strauss,
\newblock Physics of Fluids {\bf 19}, 134 (1976).

\bibitem{dahlberg1986}
Jill P. Dahlberg, David Montgomery, Gary D. Doolen and William H. Matthaeus
\newblock Journal of Plasma Physics {\bf 325}, 1 (1986).

\bibitem{zank2009}
G. P. Zank and W. H. Matthaeus,
\newblock Journal of Plasma Physics {\bf 48}, 85 (2009).

\bibitem{oughton2017}
S. Oughton, W. H. Matthaeus, and P. Dmitruk,
\newblock Astrophysical Journal {\bf 839}, 1 (2017).

\bibitem{matt1990}
W. H. Matthaeus, Melvyn L. Goldstein and D. Arron Roberts
\newblock Journal of Geophysical Research {\bf 95}, 20,673 (1990).

\bibitem{zoccoetal}
A.~Zocco and A.~A. Schekochihin,
\newblock Physics of Plasmas {\bf 18}, 102309 (2011).

\bibitem{cantwell}
Brian J. Cantwell,
\newblock Introduction to Symmetry Analysis, Cambridge University Press (2002).

\bibitem{olver}
Peter J. Olver,
\newblock Applications of Lie Groups to Differential Equations, Springer Verlag (1993).

\bibitem{CHM}
Akira Hasegawa and Kunioki Mima,
\newblock Physics Review Letters {\bf 39}, 205 (1977).

\bibitem{HK2008}
Mahouton Hounkonnou and M. M. Kabir,
\newblock Int. J. Contemp. Sciences {\bf 3}, 145 (2008).

\bibitem{WHL2018}
R. L. White, R. D. Hazeltine and N. F. Loureiro,
\newblock J. Plasma Phys. {\bf 84}, 905840204 (2018).
\end{thebibliography}
\end{document}